\documentstyle[preprint,aps]{revtex}

\def\ttbar{t\overline t}
\def\tbbar{t\overline b}
\def\bbbar{b\overline b}

\def\klcc{\kappa_L^{CC}}
\def\klnc{\kappa_L^{NC}}
\def\krcc{\kappa_R^{CC}}
\def\krnc{\kappa_R^{NC}}

\begin{document}

\draft

\preprint{$
\begin{array}{l}
\mbox{AMES-HET-96-04}\\[-3mm]
\mbox{October 1996 (revised)}\\
\end{array}
$}

\title{Unitarity Constraints on Anomalous Top Quark Couplings \\
to Weak Gauge Bosons}

\author{M. Hosch, K. Whisnant, and Bing-Lin Young}

\address{Department of Physics and Astronomy, Iowa State University,
Ames, IA 50011, USA}
\maketitle

\begin{abstract}

If there is new physics associated with the top quark, it could show up
as anomalous couplings of the top quark to weak gauge bosons, such as
$Z\ttbar$ and $W\tbbar$ vector and axial-vector couplings. We use the
processes $\ttbar\to Z^0Z^0$, $\ttbar\to W^+W^-$, and $\ttbar\to Z^0H$
to obtain the unitarity constraints on these anomalous couplings, and
combine these constraints with those from precision electroweak
data. The unitarity constraints can impose additional limits on the
anomalous couplings when the scale of new physics is as low as 2~TeV. A
nonzero measurement of such an anomalous coupling leads to an upper
limit on the new physics scale from the unitarity condition.

\end{abstract}
              
\narrowtext

\vfill
\eject

The combined CDF \cite{CDF} and D0 \cite{D0} measurements give a top
mass of $m_t = 175 \pm 9$~GeV. The large size of the top quark mass,
near the scale of electroweak symmetry breaking, suggests that the
interactions of the top quark may provide clues to the physics of
electroweak symmetry breaking and possibly evidence for physics beyond
the standard model.

If the new physics occurs above the electroweak symmetry breaking scale,
its effects can be expressed as non-standard terms in an effective
Lagrangian describing the physics at or below the new physics scale.
Such non-standard interactions, in the form of anomalous vector and
axial-vector couplings of the top quark to the $W$ and $Z$ bosons, will
affect $Z$ decay widths. The recent measurement of $R_b$ \cite{rb}, the
ratio of the decay widths $Z\rightarrow \bbbar$ and $Z\rightarrow$
hadrons, provides a motivation for studying these anomalous
couplings \cite{Z,anom}. Limits on these couplings from precision
measurements at LEP and SLC have been obtained at the one-loop level
\cite{DVMY}.

In the standard model, the gauge symmetry enforces perturbative
unitarity at all scales. In an effective theory with anomalous couplings
the gauge symmetry is explicitly broken, the renormalizability is
spoiled, and partial wave unitarity will be violated at high energies
\cite{lee}. When such anomalous couplings are present in the effective
interaction, a renormalizable Lagrangian containing the new physics
should replace the effective Lagrangian at a scale $\Lambda$ which is
below the scale where unitarity is violated, so that perturbative
unitarity is restored. Such unitarity constraints have been used
recently to put limits on anomalous Yukawa couplings of the top quark
\cite{WYZ} and on the couplings of higher dimension operators not
accessible to existing accelerators \cite{Gounaris}.

In this paper we will examine the unitarity constraints on anomalous
vector and axial vector couplings of the top quark to the $W$ and $Z$ by
calculating the amplitudes for the processes $\ttbar\to\ttbar$,
$\ttbar\to Z^0_LZ^0_L$, $\ttbar\to W^+_LW^-_L$, and $\ttbar\to Z^0_LH$,
where the $L$ subscript refers to the longitudinal component.
We parametrize the anomalous contributions to the $t {\overline t} Z$
and the $t b W$ couplings as \cite{Z,DVMY}
\begin{eqnarray}
\delta {\cal L}_{eff} &=& {-i g_2 \over 2 \cos{\theta}}
{\overline t} \gamma^\mu \left[ \klnc (1 - \gamma_5) + \krnc
(1 + \gamma_5) \right] t Z_{\mu}
\nonumber
\\
&& - {i g_2 \over {2 \sqrt2}}
\left[ {\overline t}  \gamma^\mu (\klcc (1 - \gamma_5) + \krcc
(1 + \gamma_5)) b W^+_{\mu} + h.c. \right],
\label{eq:dLag}
\end{eqnarray}
where the $\kappa$'s are dimensionless parameters which are absent in the
standard model.

The anomalous effective Lagrangian in Eq.~\ref{eq:dLag} is the simplest
possible modification of the weak sector which obeys certain elementary
constraints. It contains only dimension 4 operators, and includes only
standard neutral-current couplings for the $b$ quark at tree level. This
latter requirement arises from the experimental fact that the $b$-quark
couplings to the $Z$ are quite close to their Standard Model values. We
do not include any extra sources of CP violation, so that the $\kappa$'s
must be real. We do not include nonstandard photon couplings, so that
electrodynamics is unaltered. Finally, we note that because we have
not included all possible anomalous couplings, but rely upon the
assumption that contributions from different couplings do not cancel
each other, the results obtained here are not absolute predictions.

The anomalous coupling parameters in Eq.~\ref{eq:dLag} will affect the
interpretation of the electroweak measurements at LEP and SLC.  In
Ref.~\cite{DVMY} the authors place restrictions on the anomalous
coupling parameters using complete expressions for electroweak
measurables which even include terms which are not enhanced by
$M_t^2/M_W^2$. In order to see the effect of the unitarity constraints,
we will first update the limits with a recent analysis of the
precision electroweak data \cite{Lang}, supplemented by the latest
measurements of $R_b=\Gamma(\bbbar)/\Gamma({\rm hadrons})$ \cite{rb},
and include terms quadratic in the $\kappa$'s ignored in earlier
analyses. We find that these seemingly small quadratic terms have a significant
effect on the allowed regions in some cases. Then we will combine these
results with constraints from unitarity. We find that the unitarity
constraints can place additional limits on the anomalous couplings when the
scale of new physics is as low as 2~TeV. Furthermore, since the data is
not consistent with the Standard Model at the 90\%~CL, unitarity
constraints place an upper limit on the scale of new physics represented
by the $\kappa$'s in Eq.~\ref{eq:dLag}.

New physics at LEP and SLC can be parametrized in terms of the four
parameters $S$, $T$, $U$ \cite{stu} and $\delta_{\bbbar}$
\cite{Lang,dbb} defined by
$\Gamma(\bbbar)=\Gamma(\bbbar)_{SM}(1+\delta_{\bbbar})$, which can be
expressed in terms of $R_b$ as
$\delta_{\bbbar}=(R_b/R_b^{SM}-1)/(1-R_b)$.  The anomalous coupling
contributions to these variables (not counting the effects of a standard
model top quark and Higg boson) are \cite{DVMY}
\begin{eqnarray}
S &=& {2\over3\pi}
\left[2\krnc - \klnc - 3(\klnc)^2 -3(\krnc)^2 \right]
\log{\mu^2 \over M_Z^2},
\label{eq:Sth}
\\
T &=& {3 \over 8 \pi s_Z^2} {M_t^2 \over M_W^2}
\left[2\klcc + 4 \krnc - 4 \klnc + (\klcc)^2
+ (\krcc)^2 - 4 (\klnc-\krnc)^2 \right]
\log{\mu^2 \over M_Z^2},
\label{eq:Tth}
\\
U &=& {1 \over \pi}
\left[ - 2\klcc + 2\klnc - (\klcc)^2 - (\krcc)^2
+ 2 (\klnc)^2 + 2 (\krnc)^2 \right]
\log{\mu^2 \over M_Z^2},
\label{eq:Uth}
\\
\delta_{\bbbar} &=& {\alpha \over 9\pi}
{1 \over (8 - 12 s_Z^2 + 9 s_Z^4)} {27 s_Z^2 \over 8 (1-s_Z^2)}
{M_t^2 \over M_W^2} \log{\mu^2 \over M_Z^2}
\nonumber \\
&&\times \left[(4 s_Z^4 - 18 s_Z^2 + 9)
\left( 2\klcc +(\klcc)^2 - 4 (\krnc-\klnc)^2 \right)\right.
\nonumber \\
&&~~~\left.+ 4 (4 s_Z^4 + 2 s_Z^2 - 3) \klnc + 2 (4 s_Z^4
- 28 s_Z^2 + 15) \krnc - (28 s_Z^4 + 2 s_Z^2 - 9)(\krcc)^2 \right],
\label{eq:dbbth}
\end{eqnarray}
where $\mu$ is the renormalization scale and $s_Z^2=\sin^2\theta_W(M_Z)$.
We keep the terms quadratic in the $\kappa$'s in our analysis, as they
will affect the result even when the $\kappa$'s are not large. As in
Ref.~\cite{DVMY}, we choose the scale $\mu=2m_t$, which assumes that the
new physics is related to the top quark mass, and take $m_t=175$~GeV and
$s_Z^2=0.2311$. We have also investigated the case $\mu=m_t$ and will
comment on it later.

Recent data from LEP and SLC imply the following constraints due to
new physics contributions \cite{rb,Lang}
\begin{eqnarray}
S &=& -0.28\pm0.19^{-0.08}_{+0.17}
\label{eq:Sexp}
\\
T &=& -0.20\pm0.26^{+0.17}_{-0.12}
\label{eq:Texp}
\\
U &=& -0.31\pm0.54
\label{eq:Uexp}
\\
\delta_{\bbbar} &=& {\phantom-}0.0130\pm0.0067,
\label{eq:dbbexp}
\end{eqnarray}
where the second error in $S$ and $T$ is from varying $M_H$, the
standard model Higgs mass, between 60~GeV (the lower value) and 1~TeV
(the upper value), with a central
value at 300~GeV. In all of our calculations we take $M_H=300$~GeV. The
values for the $S$, $T$ and $U$ parameters are
taken from a recent global fit \cite{Lang}, and $\delta_{\bbbar}$ is the
latest world average \cite{rb}. The error in $\delta_{\bbbar}$ includes
the theoretical error in the Standard Model value of $R_b$, added in
quadrature with the experimental error.

The CLEO measurement of $b \rightarrow s\gamma$ \cite{cleo} also puts a
constraint on $\krcc$ \cite{krcc}. We have updated this limit using the
most recent experimental data on $m_t$ and $b\rightarrow s\gamma$ to get
\begin{equation}
-0.03 < \krcc < 0.00.
\label{eq:krcc}
\end{equation}
We note that $\krcc$ is constrained to be very small and enters into
Eqs.~\ref{eq:Sth}-\ref{eq:dbbth} only quadratically. As will be
demonstrated later, the unitarity constraints also depend on $\krcc$
only quadratically, so that we will be able to consistently ignore its
effects.

The constraints on the $\kappa$'s due to the precision electroweak data
are most easily seen by looking at the allowed regions when two of the
parameters are varied simultaneously. Figures~1a and 1b show the 90\%~CL
bounds on $\klnc$ versus $\krnc$ and $\klcc$ versus $\klnc$,
respectively, where the other parameters are set to zero in each case,
and we have set the renormalization scale at $\mu=2m_t$. The region
allowed by all the data is denoted by bold lines. As is evident
from the figures, the most important constraints on the allowed regions
are from the limits on $T$ and $\delta_{\bbbar}$, although a further
deviation from zero of the $S$ parameter would have a significant effect
on the allowed regions. The results of a three-parameter fit are
shown in Fig.~2, where allowed regions of $\klnc$ versus $\krnc$ are
shown for various values of $\klcc$.

A brief remark on the importance of the terms quadratic in the
$\kappa$'s in Eqs.~\ref{eq:Sth}-\ref{eq:dbbth} is in order. If the
quadratic terms were not included, the allowed region in Fig.~1a would
be reduced by roughly 90\% in Fig.~1a, and the allowed region in Fig.~1b is
nearly unaffected. Thus including the quadratic terms can be very
important in determining the proper allowed region even when the
magnitudes of the $\kappa$'s are less than unity. Furthermore, there is
another allowed region in the case of Fig.~1b which is a reflection of
the curves shown about $\klcc=-1$. This extra solution, which requires
the inclusion of the quadratic terms, simply changes the sign of the
$Wt\bar b$ coupling from its standard model value.


In the calculation of the unitarity constraints, we initially consider
the tree-level processes $\ttbar\rightarrow\ttbar$, $\ttbar\rightarrow
Z_L Z_L$, $\ttbar\rightarrow W_L^+ W_L^-$, and $\ttbar\rightarrow Z_L H$,
which will be affected by the anomalous couplings of Eq.~\ref{eq:dLag}.
For each reaction we consider all helicity combinations for the $t$ and
$\bar t$. The reactions with transverse vector bosons may be ignored
since their rates are suppressed in comparison with the processes
involving longitudinal vector bosons. We are most concerned with
amplitudes that grow with increasing center of mass energy, $\sqrt{s}$,
as they are guaranteed to violate unitarity at some scale. This
consideration leads us to discard the processes
$\ttbar\rightarrow\ttbar$ and $\ttbar\rightarrow Z_L H$, as they do not
grow with $\sqrt{s}$, even with nonzero anomalous couplings. For the
same reason, by considering separately each possible combination of the
top quark and the antitop quark helicity, we are led to discard the
processes $t_+ \overline{t}_- \rightarrow Z_L Z_L$, and $t_+
\overline{t}_- \rightarrow Z_L Z_L$.  This leaves four independent
processes which grow with $\sqrt{s}$. These four processes
are sufficient to constrain the anomalous couplings in our model.


For the process $\ttbar\rightarrow Z_L Z_L$, the diagrams which
contribute are the $t$- and $u$-channel exchange of a virtual top quark
and the $s$-channel Higgs exchange. To leading order in $s$ the helicity
amplitudes are
\begin{equation}
T_{++}(\ttbar \rightarrow Z_L Z_L)
= -T_{--}(\ttbar \rightarrow Z_L Z_L)
= i \sqrt2 G_f m_t \sqrt{s} [(1+\klnc-\krnc)^2-1].
\label{eq:ttZZ}
\end{equation}
For the process $\ttbar\rightarrow W_L^+ W_L^-$ the diagrams which
contribute are the $t$-channel exchange of a virtual $b$ quark, and the
$s$-channel exchange of the $Z$ boson, Higgs boson, and photon. After
retaining only leading terms proportional to $s$ and $\sqrt{s}$, the
helicity amplitudes are
\begin{eqnarray}
T_{++}(\ttbar \rightarrow W_L^+ W_L^-)
&=& -T_{--}(\ttbar \rightarrow W_L^+ W_L^-)
\nonumber
\\
&=& i \sqrt2 G_f m_t \sqrt{s} [(\klcc)^2 + 2 \klcc + (\krcc)^2
\nonumber
\\
&& +\cos(\theta) ((\klcc)^2 + 2 \klcc + (\krcc)^2 - \klnc - \krnc)]
\label{eq:tt++WW}
\\
T_{+-}(\ttbar \rightarrow W_L^+ W_L^-)
&=& i \sqrt2 G_f s \sin(\theta)[(\krcc)^2 - \krnc]
\label{eq:tt+-WW}
\\
T_{-+}(\ttbar \rightarrow W_L^+ W_L^-)
&=& i \sqrt2 G_f s \sin(\theta) [(\klcc)^2 +2 \klcc - \klnc].
\label{eq:tt-+WW}
\end{eqnarray}
As stated earlier, $\krcc$ appears only quadratically in the unitarity
conditions (as was the case for its contributions to $S$, $T$, $U$,
and $\delta_{b\bar b}$), so that its effects can be safely ignored in
view of the constraint in Eq.~\ref{eq:krcc}. Henceforth, in all of our
calculations we take $\krcc=0$.

With these expressions for the amplitudes, we may determine the
constraints from partial wave unitarity. The $J$th partial wave
amplitude for a process with helicity amplitude T is
\begin{equation}
a^J_{m,m^\prime} = {1 \over {32 \pi}} \int_{-1}^1 d(\cos\theta)
~~d^J_{m,m^\prime}(\theta)~~T_{m,m^\prime},
\label{eq:a0}
\end{equation}
where $d^J_{m,m^\prime}(\theta)$ is the Wigner $d$-function. For the
channels we are considering, $m=1$ for
$t_+ \overline t_-$, $m=0$ for $t_+ \overline t_+$, $t_- \overline t_-$,
$W^+_L W^-_L$ and $Z_LZ_L$, and $m=-1$ for $t_- \overline t_+$ (likewise
for $m^\prime$) \cite{pw}. Partial wave unitarity implies that
$\mid a^J_{m,m^\prime} \mid < 1$ for each amplitude listed in
Eqs.~\ref{eq:ttZZ}-\ref{eq:tt-+WW}.
However, the most restrictive bound comes from eigenvalues of the
coupled channel matrix for each value of $J$, each of which must also
be less than 1. We consider only the $J = 0$ and $J = 1$ partial waves,
as they give the strongest constraints. If we write the channels in the
order $t_+ \overline{t}_+$, $t_- \overline{t}_-$, $W^+_L W^-_L$, and
$Z_L Z_L$ then the coupled channel matrix for the color singlet
$J=0$ partial wave is
\begin{equation}
a^0 = {\sqrt6 G_F\over 16\pi}s \pmatrix{0 & 0 & T_1 & {T_2 \over \sqrt2} \cr
0 & 0 & -T_1 & -{T_2 \over \sqrt2} \cr
T_1 & -T_1 & 0 & 0 \cr
{T_2 \over \sqrt2} & -{T_2 \over \sqrt2} & 0 & 0},
\label{eq:coup1}
\end{equation}
where
\begin{eqnarray}
T_1 &=& {m_t \over \sqrt{s}} [(\klcc)^2 + 2\klcc +(\krcc)^2],
\label{eq:T1}
\\
T_2 &=& {m_t \over \sqrt{s}} [(1 + \klnc - \krnc)^2 -1],
\label{eq:T2}
\end{eqnarray}
and we have retained only the terms which grow with $s$.
For the color singlet $J = 1$ partial wave, the coupled channel matrix
for the channels $t_+ \overline{t}_+$, $t_+ \overline{t}_-$,
$t_- \overline{t}_+$, $t_- \overline{t}_-$, and $W^+_L W^-_L$ is
\begin{equation}
a^1 = {\sqrt6 G_F\over 48\pi}s \pmatrix{0 & 0 & 0 & 0 & T_3 \cr
0 & 0 & 0 & 0 & -\sqrt2 T_4 \cr
0 & 0 & 0 & 0 & -\sqrt2 T_5 \cr
0 & 0 & 0 & 0 & -T_3 \cr
T_3 &  -\sqrt2 T_4 & -\sqrt2 T_5 & -T_3 & 0},
\label{eq:coup2}
\end{equation}
where
\begin{eqnarray}
T_3 &=& {m_t \over \sqrt{s}} [(\klcc)^2 + 2\klcc +(\krcc)^2
- \klnc - \krnc],
\label{eq:T3}
\\
T_4 &=& (\krcc)^2 - \krnc,
\label{eq:T4}
\\
T_5 &=& (\klcc)^2 + 2 \klcc - \klnc,
\label{eq:T5}
\end{eqnarray}
and again we have retained only the terms which grow with $s$.

The characteristic equations for the roots of Eqs.~\ref{eq:coup1} and
\ref{eq:coup2} are easily found. The strongest constraint in each case
comes from the largest eigenvalue,
\begin{equation}
a^0_{max} =
{\sqrt6 G_F s \over 16\pi} \sqrt{ 2 T_1^2 + T_2^2 },
\label{eq:a0max}
\end{equation}
for $J = 0$ and
\begin{equation}
a^1_{max} =
{\sqrt6 G_F s \over 48\pi} \sqrt{ 2 \left[ T_3^2 + T_4^2 + T_5^2 \right] },
\label{eq:a1max}
\end{equation}
for $J = 1$, where partial-wave unitarity requires $a^J_{max}<1$.
Although the importance of the higher partial waves are
generally reduced by an overall factor $2J+1$, since some of the $J=1$
amplitudes grow linearly with $s$ and the $J=0$ amplitudes grow only as
$m_t\sqrt{s}$, the $J=1$ amplitudes give the most significant
constraints for the processes we are considering. Constraints on the
parameters from these bounds are shown in Figs.~3a and 3b for $\klnc$
versus $\krnc$ and $\klcc$ versus $\klnc$, respectively, for different
values of the scale $\sqrt{s}$ where unitarity is saturated, where the
other parameters are set to zero in each case. The regions allowed by
precision electroweak data are taken from the corresponding cases in
Fig.~1.

If we assume that partial-wave unitarity is obeyed up to the energy
scale of new physics, then the unitarity bounds in Fig.~3 for a given
value of $\sqrt{s}$ can be interpreted as the limits on the anomalous
couplings when the new physics scale is equal to that value of
$\sqrt{s}$. The scale at which unitarity constraints begin to encroach
on the region allowed by the LEP and SLC data varies according to the
parameter set used. The lowest energy scales for which the unitarity
constraints place additional limits on the new physics parameters
($\sqrt{s}_{min}$) are listed in the first column of Table~\ref{t:one} for
various parameter sets.

As the new physics scale is increased, the region allowed by unitarity
shrinks towards the Standard Model point (where all the $\kappa$'s are
zero). The allowed regions in Figs.~1 and 2 (determined by the precision
electroweak data) do not include the Standard Model (as expected from the
$\delta_{\bbbar}$ measurement), and predict nonzero values for the
$\kappa$ parameters. There is then a maximum value of $\sqrt{s}$
($\sqrt{s}_{max}$) for which both the unitarity and precision
electroweak constraints are satisfied.  The quantity $\sqrt{s}_{max}$
can be interpreted as an approximate upper bound on the scale of the new
physics which is embodied in the anomalous interactions of
Eq.~\ref{eq:dLag}. Values of $\sqrt{s}_{max}$ for various parameter sets
are given in second column of Table~\ref{t:one} when the 90\%~CL LEP and
SLC data are used.

When we tighten the LEP and SLC constraints by requiring 68\%~CL
agreement with the data, only $\krnc$ versus $\klcc$ has a region of
values consistent with the data when only two parameters are allowed to
vary. Figure~4 shows this allowed region and the unitarity constraints
for various values of $\sqrt{s}$. The maximum value of $\sqrt{s}$ for
which both unitarity and the electroweak constraints are satisfied is
$\sqrt{s}_{max}=3.6$~TeV for $\mu=2m_t$. When the full three-parameter
set is considered, this increases slightly to
\begin{equation}
\sqrt{s}_{max} = 3.7 {\rm~TeV}, \qquad \krnc, \klnc, \klcc \ne 0.
\label{eq:smax}
\end{equation}

We have also examined the effect on our results of changing the
renormalization scale $\mu$. Since each of the electroweak observables
in Eqs.~\ref{eq:Sth}-\ref{eq:dbbth} are proportional to
$\log(\mu^2/M_Z^2)$, reducing $\mu$ to $m_t$ will shift the
allowed regions in Figs.~1 and 2 to larger values of the couplings,
which in turn leads to smaller values of $\sqrt{s}_{min}$
and $\sqrt{s}_{max}$ (see Table~\ref{t:one}). On the other hand, if
$\mu>2m_t$ is chosen, then the allowed regions shrink in size; however,
such larger values of the renormalization scale $\mu$ are not physically
reasonable. Therefore, the energy scales listed in Table~\ref{t:one} and
Eq.~\ref{eq:smax} for $\mu=2m_t$ represent conservative estimates.
Choosing a renormalization scale as low as $m_t$ typically reduces these
values by 30\%.

In summary, our analysis shows that unitarity contraints can impose limits
on the anomalous weak gauge couplings of the top quark beyond those
given by precision electroweak data if the new physics responsible for
these couplings appears at a scale $\Lambda$ as low as 2~TeV, as
indicated by the values of $\sqrt{s}_{min}$ in Table~\ref{t:one}.
Furthermore, if the deviation from the standard model in the precision
electroweak measurements is due to such couplings, then there is a
68\%~CL upper bound on $\Lambda$ of 3.7~TeV.

\section{Acknowledgements}

We thank Xinmin Zhang for many helpful discussions.
This work was supported in part by the U.S.~Department 
of Energy under Contract DE-FG02-94ER40817. M. Hosch was also
supported under a GAANN fellowship.
  
\vfill
\eject


\begin{table}[htb]
\centering
\caption[]{Values of $\sqrt{s}_{min}$ (the lowest energy scale for which
unitarity places additional limits on the anomalous parameters) and
$\sqrt{s}_{max}$ (the highest energy scale for which both the unitarity
and electroweak constraints are satisfied) at 90\%~CL for various
parameter sets when $\mu=2m_t$. The corresponding values for $\mu=m_t$
are given in parentheses.}
\begin{tabular}{|c|c|c|}
Nonzero parameters & $\sqrt{s}_{min}$~(TeV)~~~~~~
& $\sqrt{s}_{max}$~(TeV)~~~~~~ \\ \hline
$\klnc$, $\krnc$ & 3.4 (2.6) & 19 (13) \\ \hline
$\klnc$, $\klcc$ & 3.3 (2.1) & 13 (9) \\ \hline
$\krnc$, $\klcc$ & 1.9 (1.3) & 19 (13) \\ \hline
$\klnc$, $\krnc$, $\klcc$ & 1.6 (1.3) & 20 (14) \\
\end{tabular}
\label{t:one}
\end{table}

\vfill


\centerline{FIGURE CAPTIONS}

\bigskip
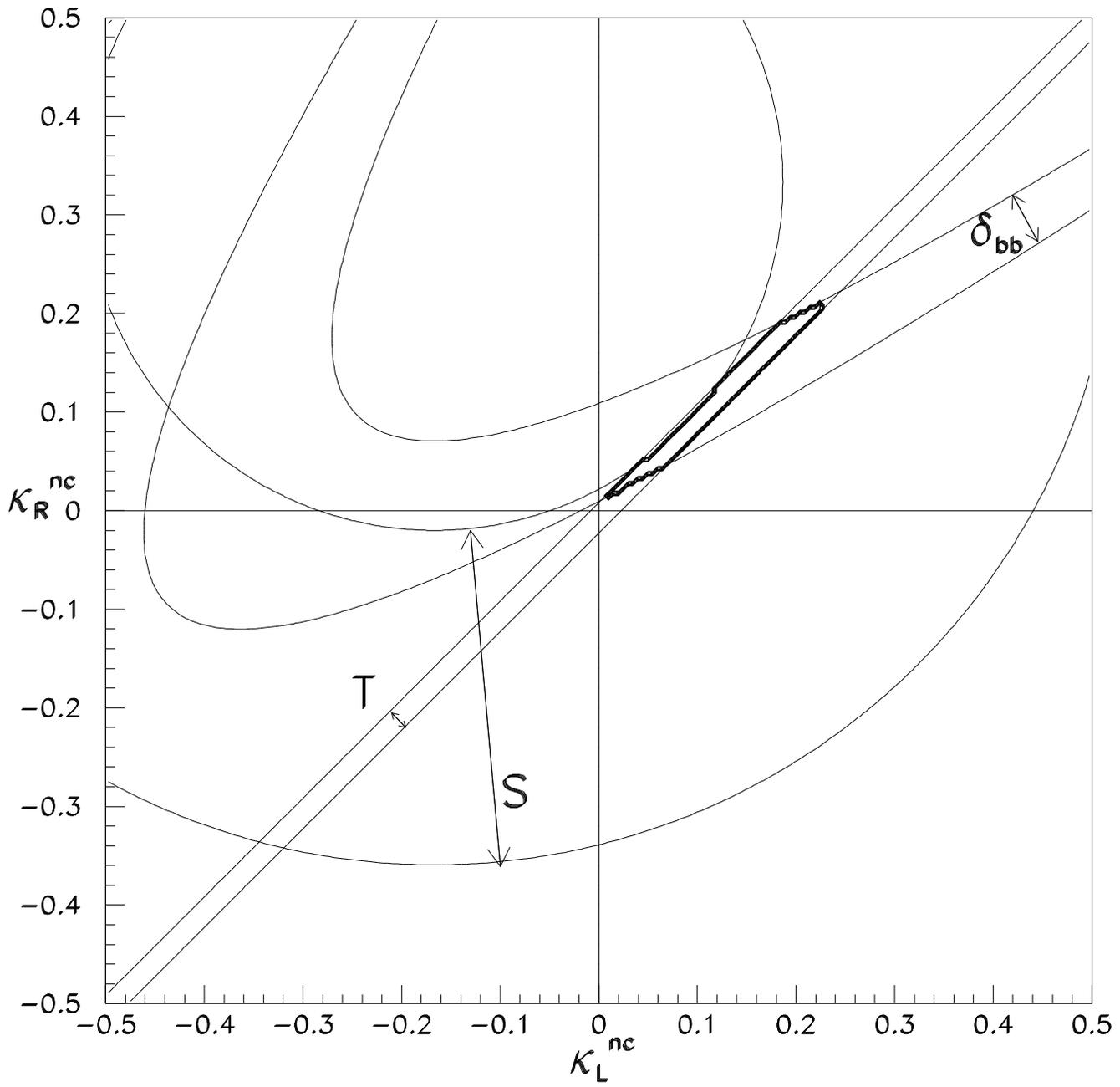
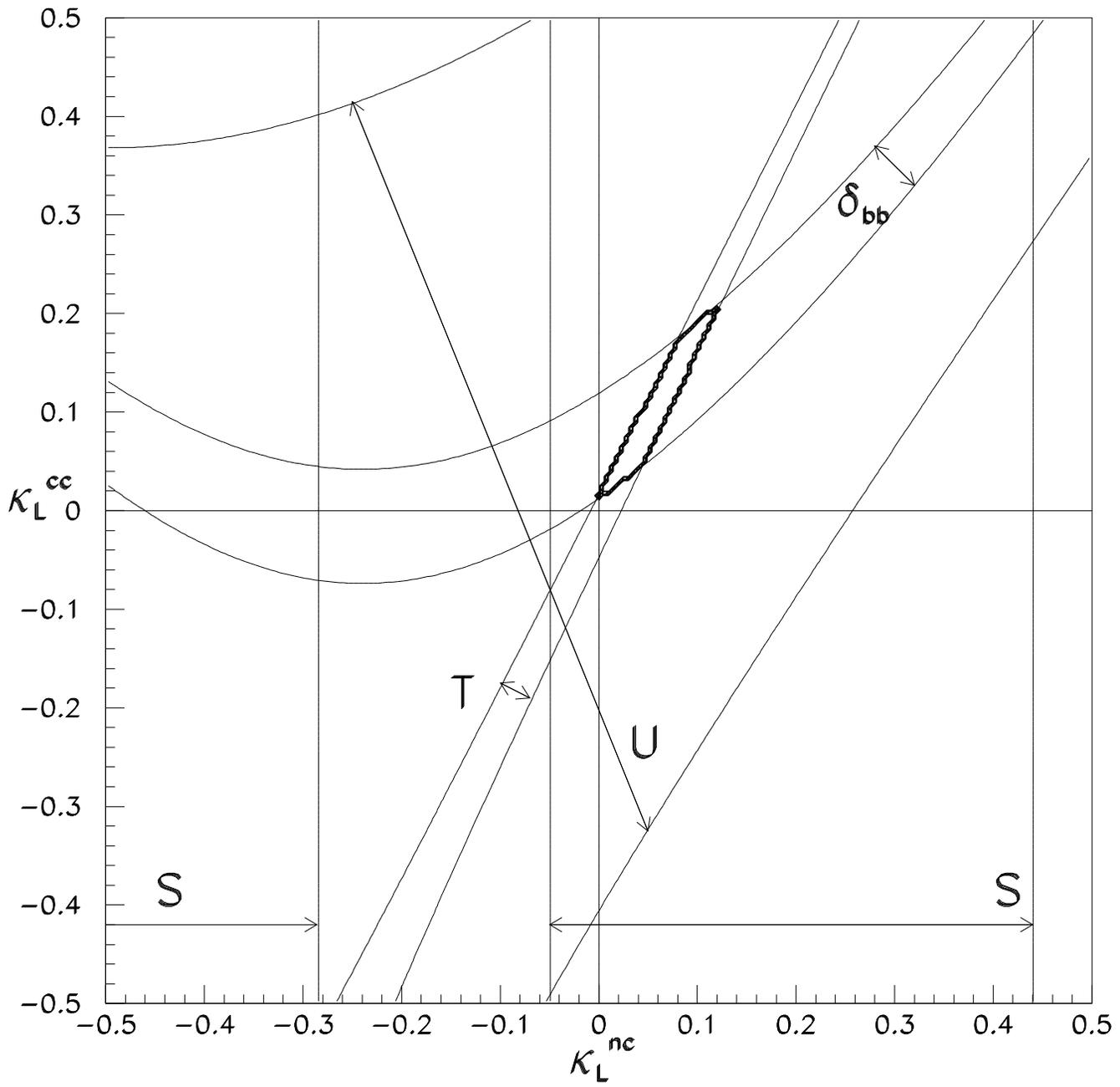
FIG.~1.
Limits from precision LEP and SLC data on (a) $\krnc$ vs. $\klnc$ for
$\klcc = \krcc = 0$, and (b) $\klcc$ vs. $\klnc$ for $\krnc = \krcc = 0$,
using the 90\% CL limits on $S$, $T$ and $U$ from Ref.~\cite{Lang}, and on
$\delta_{\bbbar}$ from Ref.~\cite{rb}. The regions allowed by the
electroweak variables $S$, $T$, $U$ and $\delta_{\bbbar}$ are indicated
by the arrows. In Fig.~1(a) the entire region shown is allowed by $U$.
In each case the region allowed by all of the electroweak data lies inside
the bold lines.

\bigskip
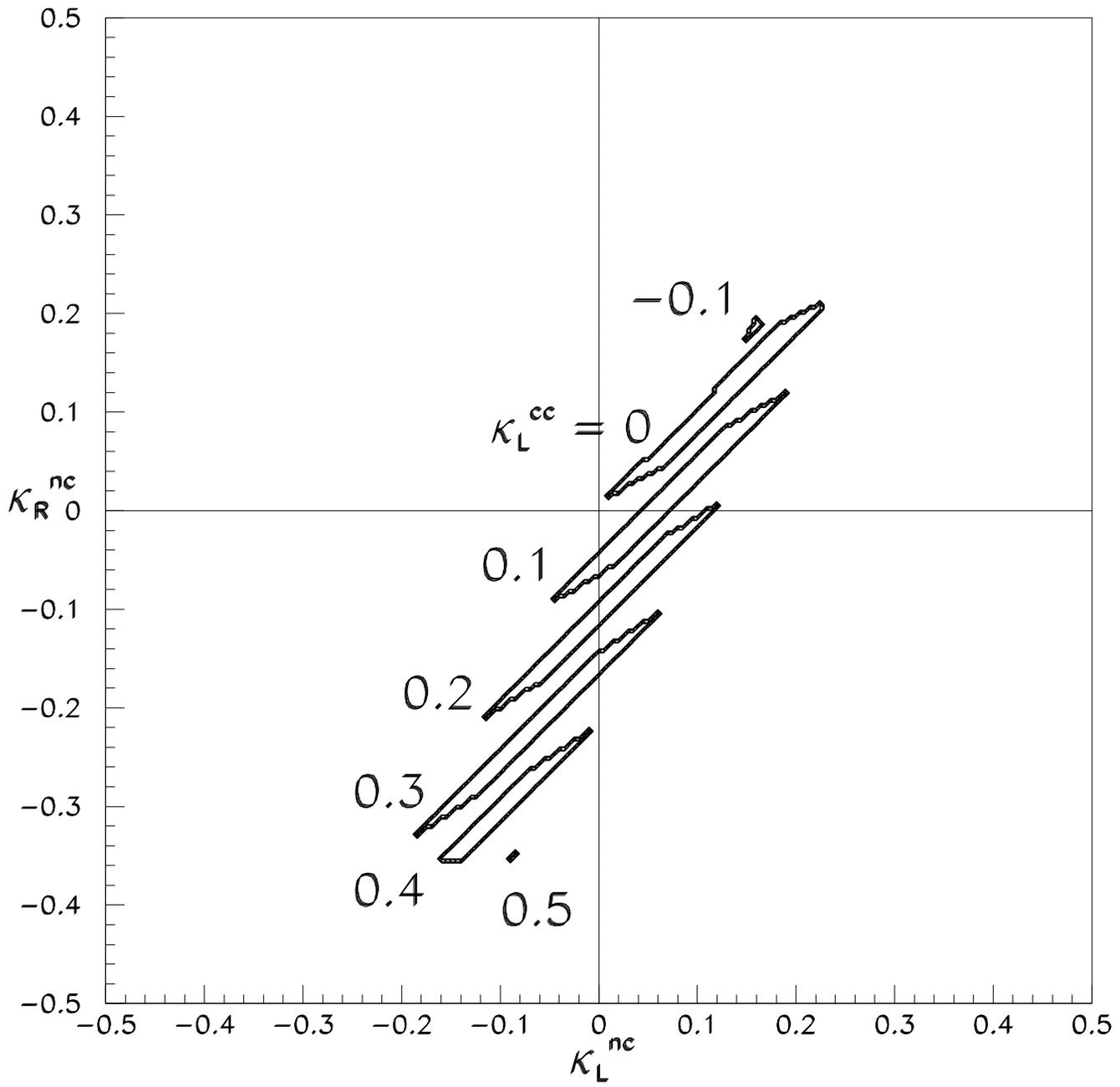
FIG.~2.
Allowed region from precision LEP and SLC data of $\klnc$ vs. $\krnc$ for
several values of $\klcc$ with $\krcc = 0$, using the 90\% CL limits
from Refs.~\cite{rb} and \cite{Lang}.

\bigskip
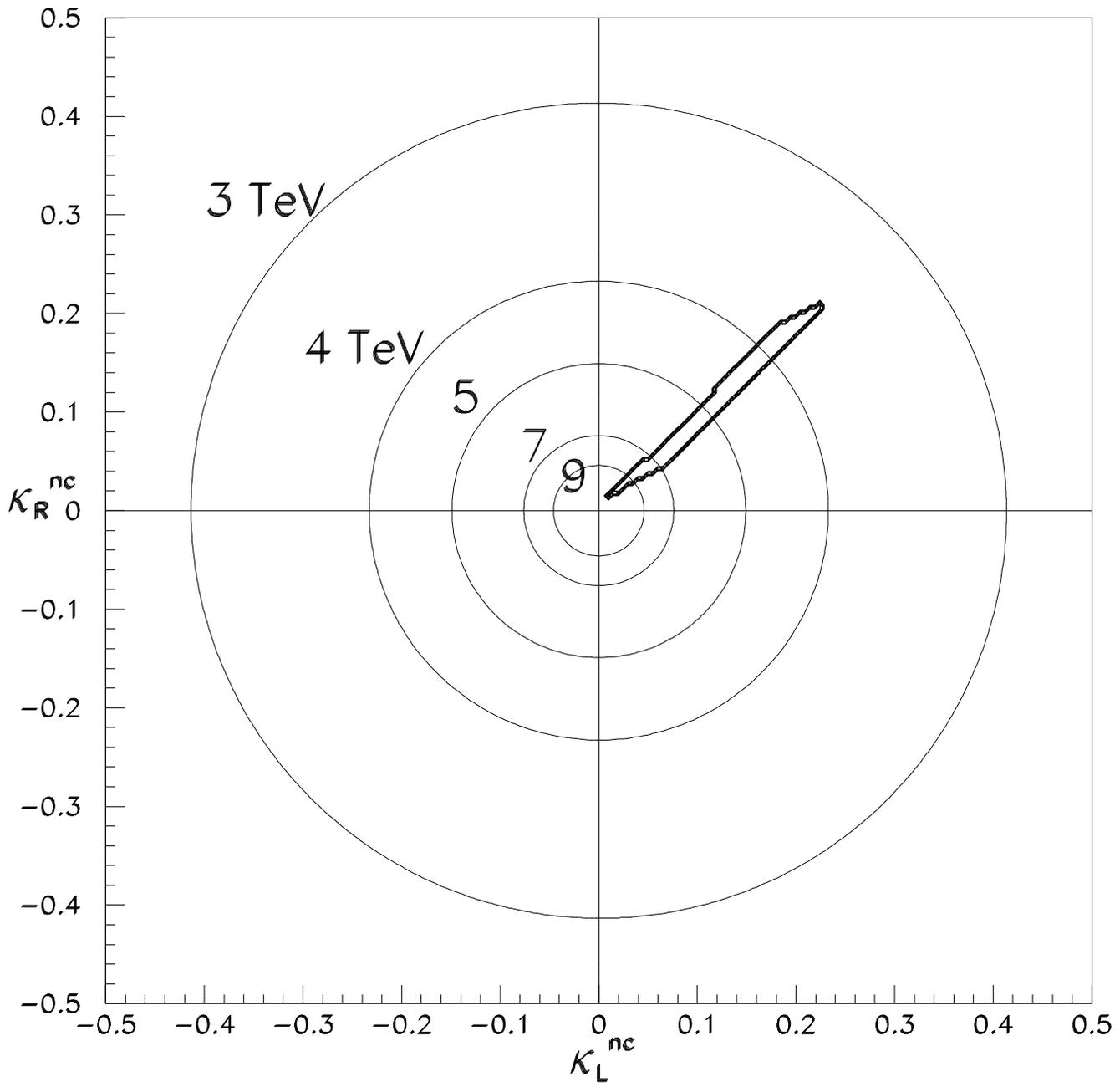
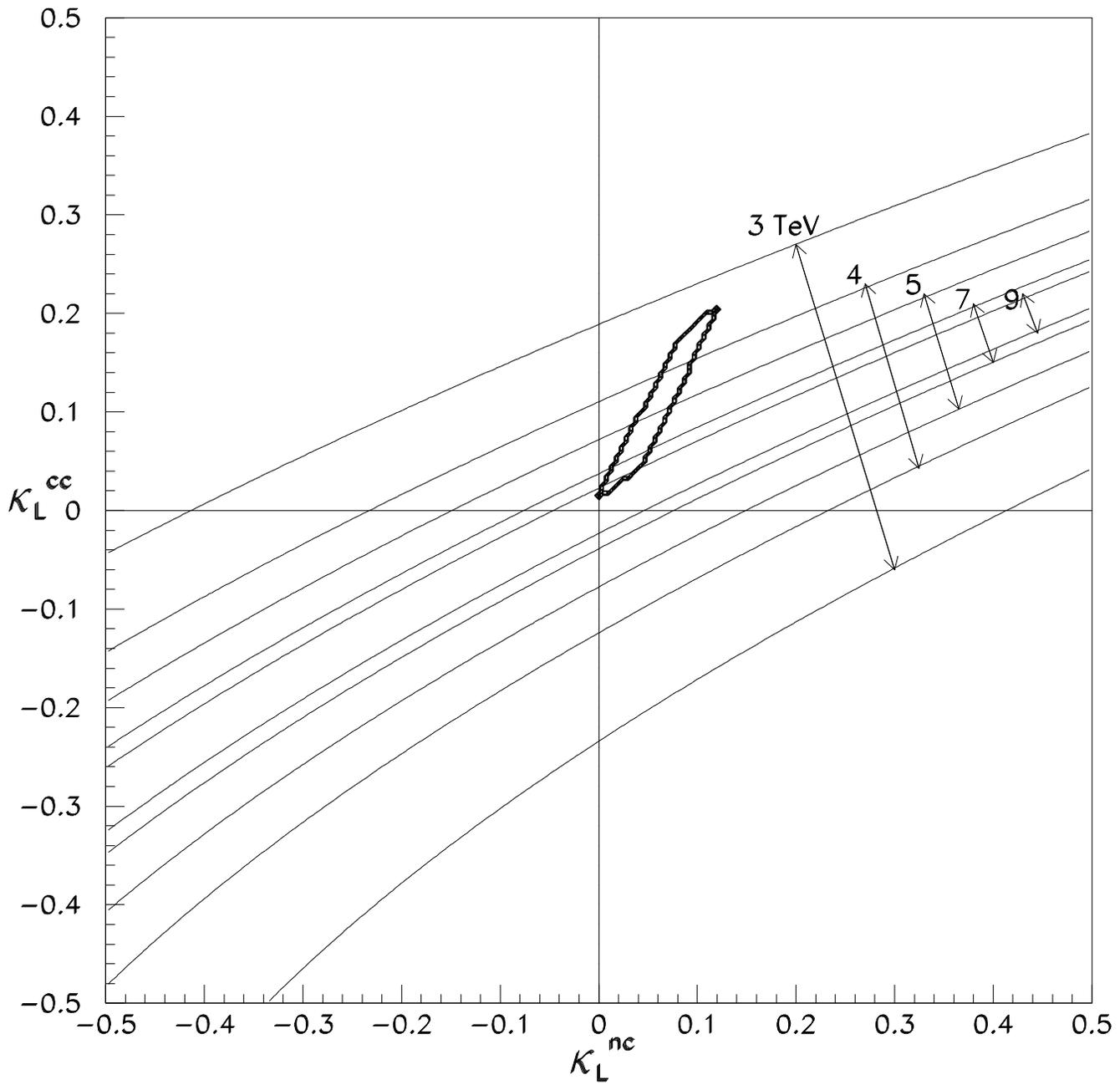
FIG.~3.
Unitarity limits on (a) $\krnc$ vs. $\klnc$ for $\klcc = \krcc = 0$, and
(b) $\klcc$ vs. $\klnc$ for $\krnc = \krcc = 0$, shown for several
values of $\sqrt{s}$. The regions allowed by unitarity lie inside the
circles in (a), and between the lines for each energy scale
in (b). The 90\%~CL regions allowed by LEP and SLC data, taken from
Fig.~1, are also shown.

\bigskip
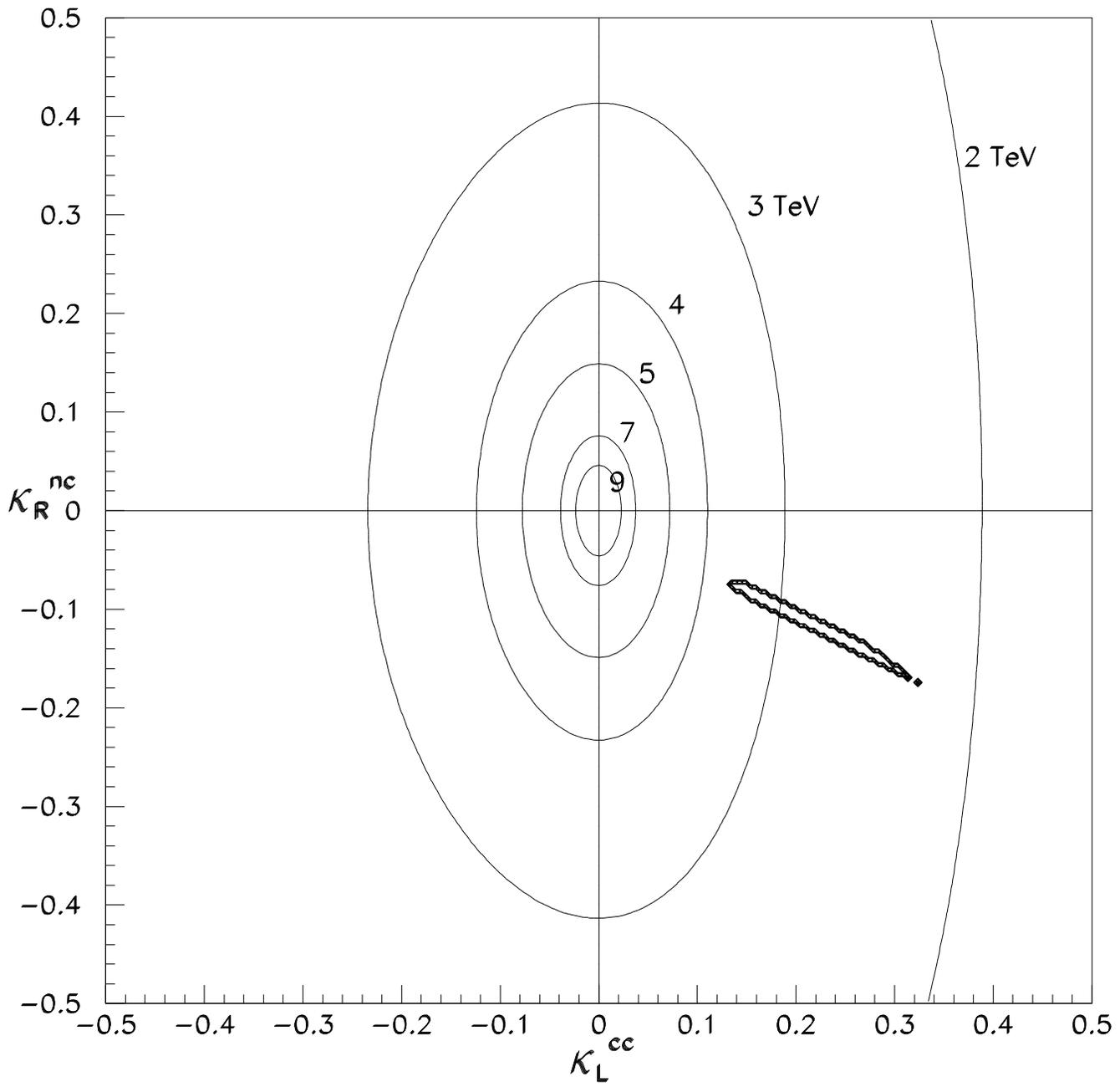
FIG.~4.
Unitarity limits on $\krnc$ vs. $\klcc$ for $\klnc=\krcc=0$, shown for
several values of $\sqrt{s}$. The regions allowed by unitarity lie
inside the ellipses for each energy scale. The 68\%~CL region allowed
by LEP and SLC data is also shown.

\end{document}